%% file: main.tex
\documentclass{article}
\usepackage{spconf,amsmath,graphicx}

\usepackage{todonotes}
\usepackage{verbatim}
\usepackage{wrapfig}
\usepackage{siunitx}
\usepackage{listings}
\usepackage{lstautogobble}
\usepackage[numbers]{natbib}
\usepackage{subfig}
\usepackage{booktabs}
\usepackage{hyperref}

\usepackage{amssymb}

\setcitestyle{square}

\usepackage{enumitem}
\setlist[itemize]{leftmargin=*} 
\setlist{nosep}

\newcommand\extrafootertext[1]{%
    \bgroup
    \renewcommand\thefootnote{\fnsymbol{footnote}}%
    \renewcommand\thempfootnote{\fnsymbol{mpfootnote}}%
    \footnotetext[0]{#1}%
    \egroup
}

\newcommand{\rough}[1]{#1} 

\title{Speech synthesis and control using differentiable DSP}

\name{Giorgio Fabbro $^{\text{1,2}}$ \qquad Vladimir Golkov $^{\text{2}}$ \qquad Thomas Kemp $^{\text{1}}$ \qquad Daniel Cremers $^{\text{2}}$}
\address{$^{\text{1}}$ Sony Europe B.V., Stuttgart \qquad $^{\text{2}}$ Computer Vision Group, Technical University of Munich}

\begin{document}
\ninept
\maketitle
\begin{abstract}
Modern text-to-speech systems are able to produce natural and high-quality speech, but speech contains factors of variation (e.g.~pitch, rhythm, loudness, timbre)\ that text alone cannot contain.
In this work we move towards a speech synthesis system that can produce diverse speech renditions of a text by allowing (but not requiring) explicit control over the various factors of variation. We
propose a new neural vocoder that offers control of such factors of variation.
This is achieved
by employing differentiable digital signal processing (DDSP) (previously used only for music rather than speech), which exposes these factors of variation.
The results show that the proposed approach can produce natural speech with realistic timbre, and individual factors of variation can be freely controlled.
\end{abstract}
\begin{keywords}
Neural vocoder, speech synthesis, digital signal processing, neural networks, deep learning
\end{keywords}
%

\section{Introduction}\label{sec:intro}
A sentence can be pronounced in many different ways: fast or slow, happy or angry, with emphasis on certain words. We can say much more than what is written
because we are able to control some fundamental aspects of the production of speech. In this work, we call these aspects \emph{factors of variation}, since their variation affects the overall meaning of speech. These factors of variation include \emph{pitch}, \emph{rhythm}, \emph{loudness}, and \emph{timbre}. 
In this work, we move towards a speech synthesis system that can produce diverse speech renditions of a text by allowing explicit control over the various factors of variation.

In literature, control is achieved in different ways.
Here we make an important distinction between models that \emph{require} control and models that offer \emph{optional} control.
Models that require control expect additional inputs to generate speech, whereas models that offer optional control disentagle the data into various components and provide the possibility to modify them.
Optional control is preferable, \rough{as not always affecting the generation is desired and not always we possess all the required inputs.}

Digital signal processing (DSP) algorithms are central methods in audio engineering.
Recently, differentiable DSP (DDSP) \cite{EngelHantrakulGuRoberts2020} has been introduced as a new method to generate audio with deep learning: DSP algorithms 
are used as parts of a neural network, ensuring end-to-end optimization. DDSP so far has only been applied to
music. We use DDSP to generate speech.
Since DDSP generators directly map the controllable variables to audio, we propose control where the audio is generated (i.e.~in the neural vocoder)\rough{, and not earlier in the pipeline as others do.}
This choice has several advantanges: it can be used not only with a spectrogram generator in text-to-speech~(TTS), but also to modify
properties of existing audio clips. \rough{Moreover, our neural vocoder provides direct control to any spectrogram generator, without the effort and the difficulty of redesigning the latter to offer
control.}
This work is intended as a first step towards \rough{DDSP-based} speech synthesis systems \rough{for a wide range of applications} that are able to let the user affect aspects of speech that text alone cannot encode.

\begin{figure}[b!]
\begin{minipage}[b]{1.0\linewidth}
  \centering
  \centerline{\includegraphics[width=8.5cm]{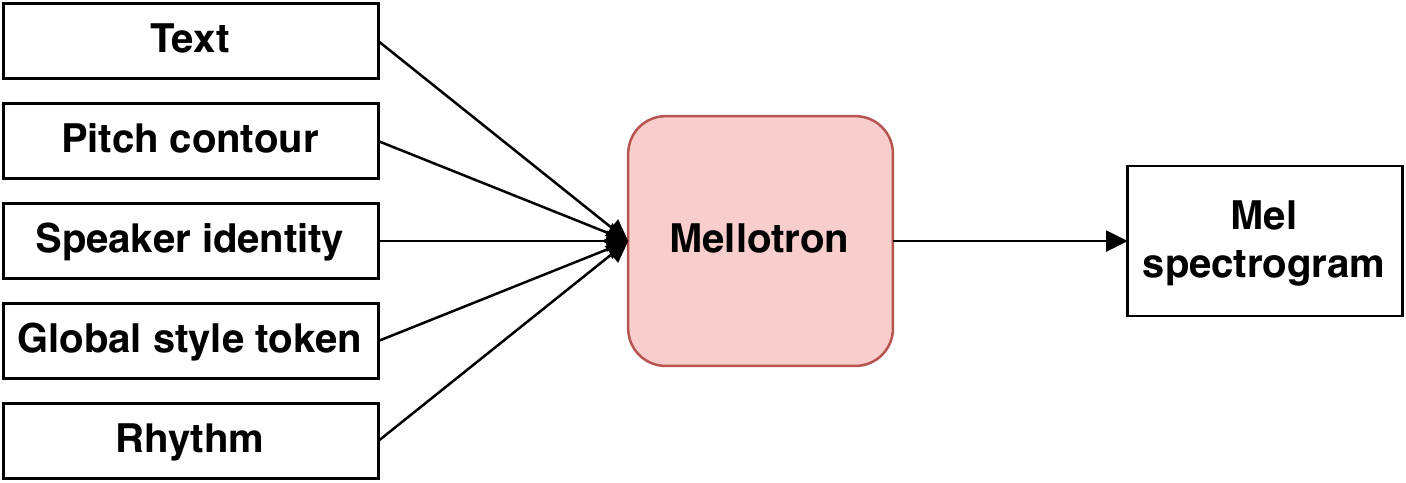}}
  \label{fig:control_comp_mellotron}
  \centerline{(a) Mellotron
  }\medskip
\end{minipage}
\begin{minipage}[b]{1.0\linewidth}
  \centering
  \centerline{\includegraphics[width=8.5cm]{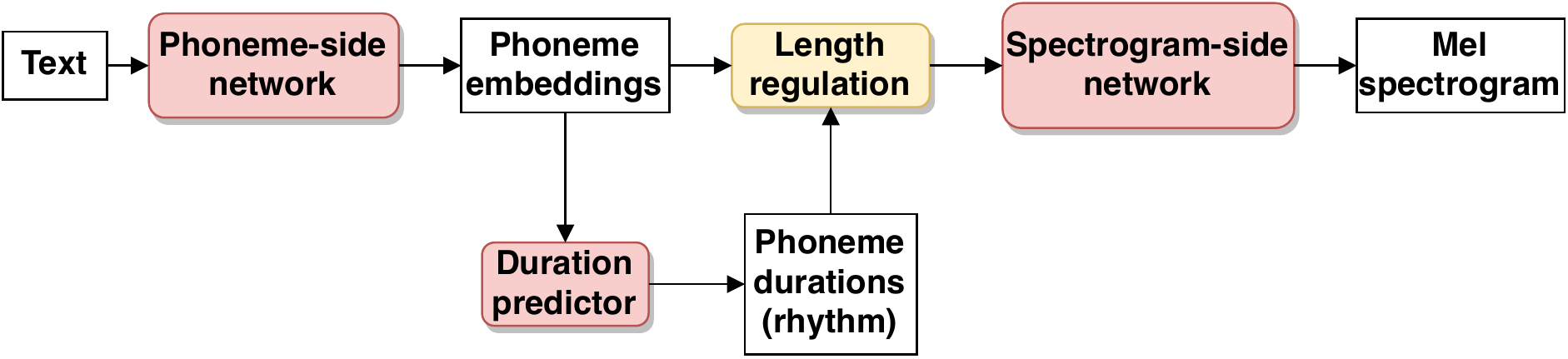}}
  \label{fig:control_comp_fastspeech}
  \centerline{(b) FastSpeech}
\end{minipage}
\caption{
Comparison of two existing models that use factors of variation such as rhythm to affect the audio 
generation.
Red blocks are learned models, yellow blocks are fixed operations.
(a) Mellotron~\cite{ValleLiPrengerCatanzaro2019} does not infer the factors of variation, it always requires them as input.
(b) FastSpeech~\cite{RenRuanTanQinZhaoZhaoLiu2019} infers phoneme durations (i.e.~rhythm) and offers (but \emph{does not require}) their control.
\rough{Offering control without requiring it makes a model applicable in a wider variety of settings.}
}
\label{fig:control_comp}
\end{figure}

\begin{figure*}[tb]

\centering
\includegraphics[width=0.67\textwidth]{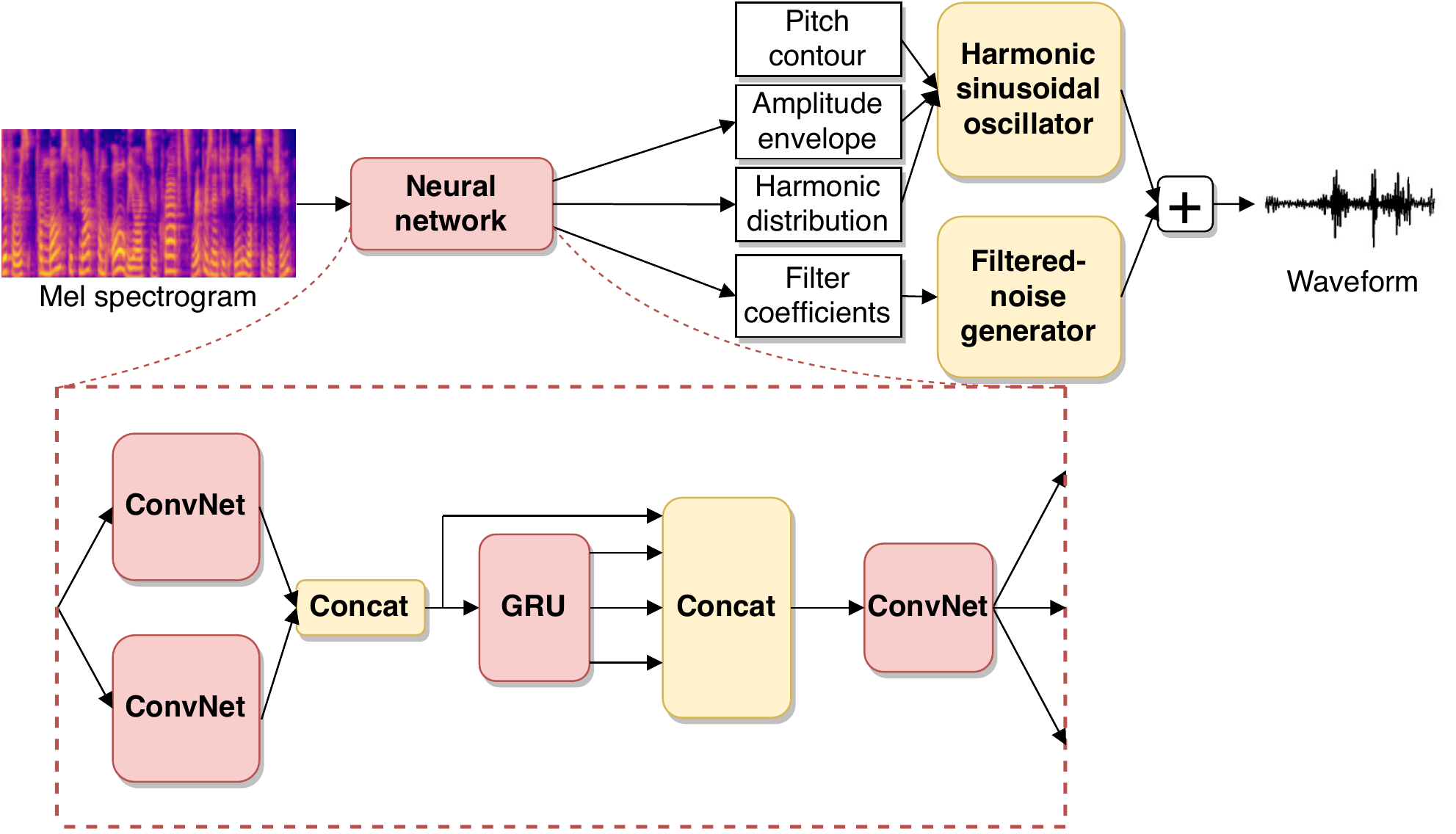}
\caption{
The proposed DDSP-based neural vocoder.
The neural network decomposes the spectrogram into control variables 
(amplitude envelope, harmonic distribution, and filter coefficients)
for the DSP generators: this allows us to control them, if needed.
We use the ground-truth pitch contour during training instead of letting the neural network infer it, in order to study in isolation the network's capabilities of
modeling the other three control variables.
The oscillator produces vowels, the noise generator produces consonants.
}
\label{fig:ddsp_architecture}
\end{figure*}

\section{Related Work}\label{sec:related_work}

The goal of TTS is to convert a text sequence into an audio rendition of someone's voice pronouncing that text. The typical TTS system employs two steps to achieve that goal. First, a model converts the text into an acoustic representation of speech, usually a mel spectrogram. We call such a model a \emph{(mel) spectrogram generator}. Then, another model converts the mel spectrogram into the audio waveform. This model is usually referred to as a \emph{(neural) vocoder}.
In existing methods, control of factors of variation happens during the mel spectrogram generation. Our method, on the other hand, offers control in the neural vocoder.
In the following we first outline some models that generate mel spectrograms while offering control. Then we outline the major neural vocoders in literature.

Various approaches have been proposed to affect the mel spectrogram generation from text. Those that \emph{require} control
use additional inputs to the networks as in Fig.~\ref{fig:control_comp}\rough{a}~\cite{WangStantonZhangSkerryRyanBattenbergShorXiaoRenJiaSaurous2018,ShechtmanRabinovitzSorinKonsHoory,ValleLiPrengerCatanzaro2019,ShenJiaChrzanowskiZhangEliasZenWu2020,BaeBaeJooLeeLeeCho2020}.
Those that offer control  (but \emph{do not require} it)  
aim at controlling latent variables while ensuring that they are interpretable \cite{RaitioRasipuramCastellani2020,ValleShihPrengerCatanzaro2020,HsuZhangWeissZenWuWangCaoJiaChenShenetal2018,HabibMariooryadShannonBattenbergSkerryRyanStantonKaoBagby2019}, or
disentangle one interpretable control variable (often the duration of the utterance) and make it controllable as in Fig.~\ref{fig:control_comp}\rough{b}~\cite{RenRuanTanQinZhaoZhaoLiu2019,RenHuTanQinZhaoZhaoLiu2020,KimKimKongYoon2020}.

Existing neural vocoders use various principles such as autoregressive networks~\cite{OordDielemanZenSimonyanVinyalsGravesKalchbrennerSeniorKavukcuoglu2016},
flows \cite{parallelwavenet,PrengerValleCatanzaro2018,KimLeeSongKimYoon2019,ZhaiGaoXueRothchildWuGonzalezKeutzer2020,PingPengZhaoSong2019}, adversarial learning \cite{DonahueMcAuleyPuckette2019,BinkowskiDonahueDielemanClarkElsenCasagrandeCoboSimonyan2019,melgan,parallelwavegan}, and source-filtering \cite{WangTakakiYamagishi2019}.
These principles do not expose the factors of variation as interpretable variables and thus do not allow their control.
We propose using a different principle to generate speech, namely DDSP \cite{EngelHantrakulGuRoberts2020}, that so far was used only for music synthesis.

\section{Methods}\label{sec:methods}
\rough{
Our objective is to have a model that offers (but \emph{does not require}) control of factors of variation, so that we can modify them if needed.
A simple way of achieving this 
is to include in the model architecture a fixed (i.e.~not learned) operation that requires such variable (and make sure that no other operation will further perturb its outcome). For instance, the length regulation algorithm in Figure \ref{fig:control_comp}b requires phonemes duration as input: in this way, the authors can disentangle the speech rhythm from the data and make it available to the user for control.
Another approach is to use DSP algorithms such as oscillators, since they generate audio signals using variables that we would like to control.
Following~\cite{EngelHantrakulGuRoberts2020}, we insert differentiable DSP algorithms to synthesize audio (in our case speech): our model architecture has then fixed operations that use our controllable variables. 
}

\subsection{Network Architecture}
Our neural vocoder is depicted in Figure~\ref{fig:ddsp_architecture}. The model takes as input the mel spectrogram of a sentence and uses a recurrent neural network to decompose it into a set of control variables. These variables are input to a harmonic oscillator and a filtered-noise generator. These two final blocks output two audio sequences, which get summed to produce the final speech signal. The oscillator and the noise generator are differentiable (as in~\cite{EngelHantrakulGuRoberts2020}), therefore we can apply a loss to the output signal and expect the gradients to flow back to the weights of the neural network. Note that the usage of an oscillator to model the voiced parts of speech (vowels) and a filtered-noise generator to model consonant sounds makes our model similar to spectral modeling synthesis~\cite{sms} and to neural source-filter models~\cite{WangTakakiYamagishi2019}.

\subsubsection{Trainable Layers}

For simplicity, we use the same network architecture as the decoder in DDSP~\cite{EngelHantrakulGuRoberts2020}. The decoder first applies to each of its inputs a 1D ConvNet with layer normalization, leaky ReLUs, and filter size~1 (equivalent to applying a multilayer perceptron to each time frame),
then concatenates the outputs of the ConvNets over the channels dimension and feeds them to a gated recurrent unit (GRU) that processes them over the time dimension. The output of the GRU is concatenated with its input and fed to another ConvNet. Its output is then split over the channels dimension to create the desired control variables.
Our network is different from the DDSP decoder~\cite{EngelHantrakulGuRoberts2020} only in the input stage: their inputs are intermediate variables (pitch, loudness, and a latent variable) each processed by a separate ConvNet, whereas our input is only the mel spectrogram. Therefore, we reduced the number of parallel ConvNets in the input stage. Still, in~\cite{EngelHantrakulGuRoberts2020} the authors could have used a single ConvNet and feed it with the three inputs concatenated along the channels dimension, but they used three separate ConvNets in parallel. 
This indicates that the network performs better if it processes different information (for example pitch and loudness) in separate ConvNets (without cross-talk) before merging the extracted features.  
Similarly, we
use two ConvNets in parallel.
In this way the network can learn to extract and process different parts of information from the spectrogram separately before merging high-level features.
Hyperparameters
are listed in Section~\ref{sec:experiments}.

In order to make the optimization easier in the first experiments and to focus on learning to infer a specific subset of control variables, we provide to the harmonic oscillator the pitch contour extracted from the original audio, so that the efforts of the neural network are focused on loudness, timbre, and rhythm. This does not affect the capabilities of the neural vocoder in controlling pitch.

We let the network work at an intermediate temporal resolution that is higher than in the mel spectrogram but lower than in the audio. To this end, the mel spectrogram is upsampled by a factor of $8.5$ 
using bilinear interpolation before going into the network. 
Lower upsampling factors yielded lower audio quality, whereas higher ones yielded similar quality but slower computation.
The network output is further upsampled using bilinear interpolation to reach the temporal resolution of the audio.

\subsubsection{Harmonic Sinusoidal Oscillator}
The time-upsampled outputs of the neural network are used to control two DSP generators: a harmonic sinusoidal oscillator and a filtered-noise generator. The harmonic sinusoidal oscillator is controlled by the fundamental frequency (pitch contour) $f_1(n) \in \mathbb{R}^+$ for each time instant $n$,
an amplitude envelope  
$A(n) \in \mathbb{R}^+$ for each $n$,  
and a distribution $c_k(n)$ over harmonics that for each time step contains the weight to be applied to each harmonic (the weights sum up to 1 for every time step). The distribution $c_k$  
characterizes the timbre and the vowel.
Following~\cite{EngelHantrakulGuRoberts2020},  
the amplitude for each harmonic is $A_k(n) = A(n)c_k(n)$.

The oscillator generates a superposition $y(n)$ of harmonically related sinusoidal signals, i.e.~signals whose
frequency $f_k(n)$  
is an integer multiple of the fundamental frequency $f_1(n)$
, i.e.~$f_k(n) = k f_1(n)$ with $k \in \mathbb{Z}^{+}$. Since digital signals are bandlimited, we only consider a finite number $H$ of harmonics, i.e.~$1 < k < H$. The output of the oscillator is:
\begin{equation}\label{eq:harm_sin_oscillator}
    y(n) = \sum_{k=1}^{H} A_k(n) \sin (\phi_k(n)),
\end{equation}
where
$\phi_k(n) = 2\pi \sum_{i=0}^n f_k(i)$ is the so-called instantaneous phase for time instant $n$ and harmonic $k$.
The only hyperparameter that must be chosen for the harmonic oscillator is the number $H$ of harmonics it generates. The dataset we used was recorded at a sampling frequency $f_s = \SI{22050}{Hz}$, so the Nyquist frequency is $f_{\mathrm{Nyq}} =  
f_s/2
= \SI{11025}{Hz}$. A generic female speech signal has a fundamental frequency with lower bound $f_1^{(\mathrm{min})} = \SI{165}{Hz}$~\cite{ingo1994principles} (we model female voice because our dataset contains only female speech). Therefore, in order to effectively model all harmonic content that our signal potentially has, we need $H =
f_{\mathrm{Nyq}} / f_1^{(\mathrm{min})}
\approx 67$ harmonics.

\subsubsection{Filtered-Noise Generator}
To add non-harmonic
components (consonants) to our synthesis process, we use a module that generates white noise and filters it with a linear time-varying filter bank~\cite{EngelHantrakulGuRoberts2020}.
We control this  
\emph{filtered-noise generator}
by letting the neural network output 
parameters for the time-varying filter bank.  
For speed,
the filtering operation occurs in the frequency domain: 
the time-upsampled outputs of the neural network are elementwise multiplied with white noise in the frequency domain.
The signal is then converted to time-domain  
using the overlap--add method   
to account for the overlap between adjacent time frames.
We can choose the frequency resolution of the filter by changing the number $M$ of frequency-domain coefficients (network output channels).
By increasing $M$, we decrease the amount of frequencies that each coefficient corresponds to,
therefore increasing the filter resolution. 
We found that $M = 101$ gave us good results, while keeping the computational load low.

\subsubsection{Training Objective}
Our loss is computed using spectrograms with different time resolutions, similarly to the multi-scale spectrogram loss 
from DDSP~\cite{EngelHantrakulGuRoberts2020}.
Since we propose to adapt DDSP to speech,
we define the loss via mel spectrograms, which are widely used for speech applications, instead of short-time Fourier transforms (STFT).
A mel spectrogram is computed by mapping the result of STFT to the mel scale. This implies that in our case the frequency resolution will remain always constant for STFT results computed with different time-frequency resolutions, as we map all of them to the same mel scale. Therefore, our loss is
\begin{equation}\label{eq:mel_spectrogram_loss}
    \mathcal{L} = \sum_{i \in R} \left\lVert S_i^{\mathrm{mel}} - \hat{S}_i^{\mathrm{mel}} \right\rVert_1 + \alpha \left\lVert \log S_i^{\mathrm{mel}} - \log \hat{S}_i^{\mathrm{mel}} \right\rVert_1,
\end{equation}
where $R = \{2048, 1024, 512, 256, 128, 64\}$ is the set of time-frequency resolutions measured in the number of samples, $S^{\mathrm{mel}}$ is the mel spectrogram of the ground-truth audio and $\hat{S}^{\mathrm{mel}}$ is the mel spectrogram of the generated audio.
In all the experiments, we used mel spectrograms with 80 frequency bins. The time overlap of the audio data between neighbouring frames in each spectrogram is~$75\%$.

\subsection{Experimental Setup}\label{sec:experiments}

We trained on
the LJSpeech dataset~\cite{ljspeech17}, which contains 13100 short audio clips of one female speaker reading non-fictional passages. The clips vary in length between 1 and 10 seconds and have an average length of 6.5 seconds. The total length of the dataset is approximately 24 hours. The audio sampling frequency is $f_s = \SI{22050}{Hz}$ and the encoding is 16-bit PCM WAV. No pre-processing was applied to the original audio and the input mel spectrograms were obtained from the ground truth audio by applying the STFT where each frame is the result of an FFT applied to 1024 samples of audio; each frame overlaps with the previous one and the next one by $75\%$. They have 80 frequency bins that cover the interval $[\SI{0}{Hz} ; \SI{8000}{Hz}]$.
We provided our model with ground truth pitch contours extracted with the YIN algorithm~\cite{yin} using the same parameters as for the mel spectrograms. 
We used the first $12822$~clips as the training set, and the remaining $278$~clips (i.e.~the files named \texttt{LJ050-*.wav}) as the test set. For validation, we randomly selected a portion of the training set.
As baselines, we used the official implementation of WaveGlow~\cite{PrengerValleCatanzaro2018} and our implementation of WaveNet~\cite{OordDielemanZenSimonyanVinyalsGravesKalchbrennerSeniorKavukcuoglu2016} in NNabla\footnote{www.github.com/sony/nnabla}.

We used the Adam optimizer~\cite{kingma2014adam} with learning rate $10^{-4}$, $\beta_1=0.9$ and $\beta_2=0.999$. We applied an exponential learning rate decay of $0.98$ every $10^4$ iterations.
We stopped training after $4 \cdot 10^5$ iterations as we could not notice any more improvement.

Our best configuration
uses the following hyperparameters: batch size 8, batch length $6 \cdot 10^4$ samples,  
and $M = 101$.
The GRU layer in the architecture has 512 units. 
Each ConvNet has 3 layers with filter size 1 and 512 filters in the hidden layers.
We also tried a smaller architecture (1 layer in each ConvNet with 256 filters in hidden layers), 
but results were worse.

\section{Results and Discussion}\label{sec:results}

\subsection{Audio Clips}
Typical clips generated by our neural vocoder are available on the project's GitHub page\footnote{\url{https://thesmith1.github.io/DDSPeech/}}. Our speech is fully intelligible and we model accurately the speaker's timbre. Some consonants still sound slightly artificial and some  slight noise from the noise generator can be heard even when vowels are pronounced.

On the same webpage, we also provide some clips to showcase the control capabilities of our neural vocoder. In particular, we show how we can freely change the pitch of the utterance, either by modification or by substitution.
Moreover, 
in some cases we perceive traces of the original pitch
(e.g.~in the clip where the pitch is modified to be constant). 
These traces appear to be encoded in the timbre.

\subsection{Synthesis Speed}\label{sec:synthesis_speed_results}
\begin{figure}[tb]

\begin{minipage}[b]{1.0\linewidth}
  \centering
  \centerline{\includegraphics[width=8.5cm]{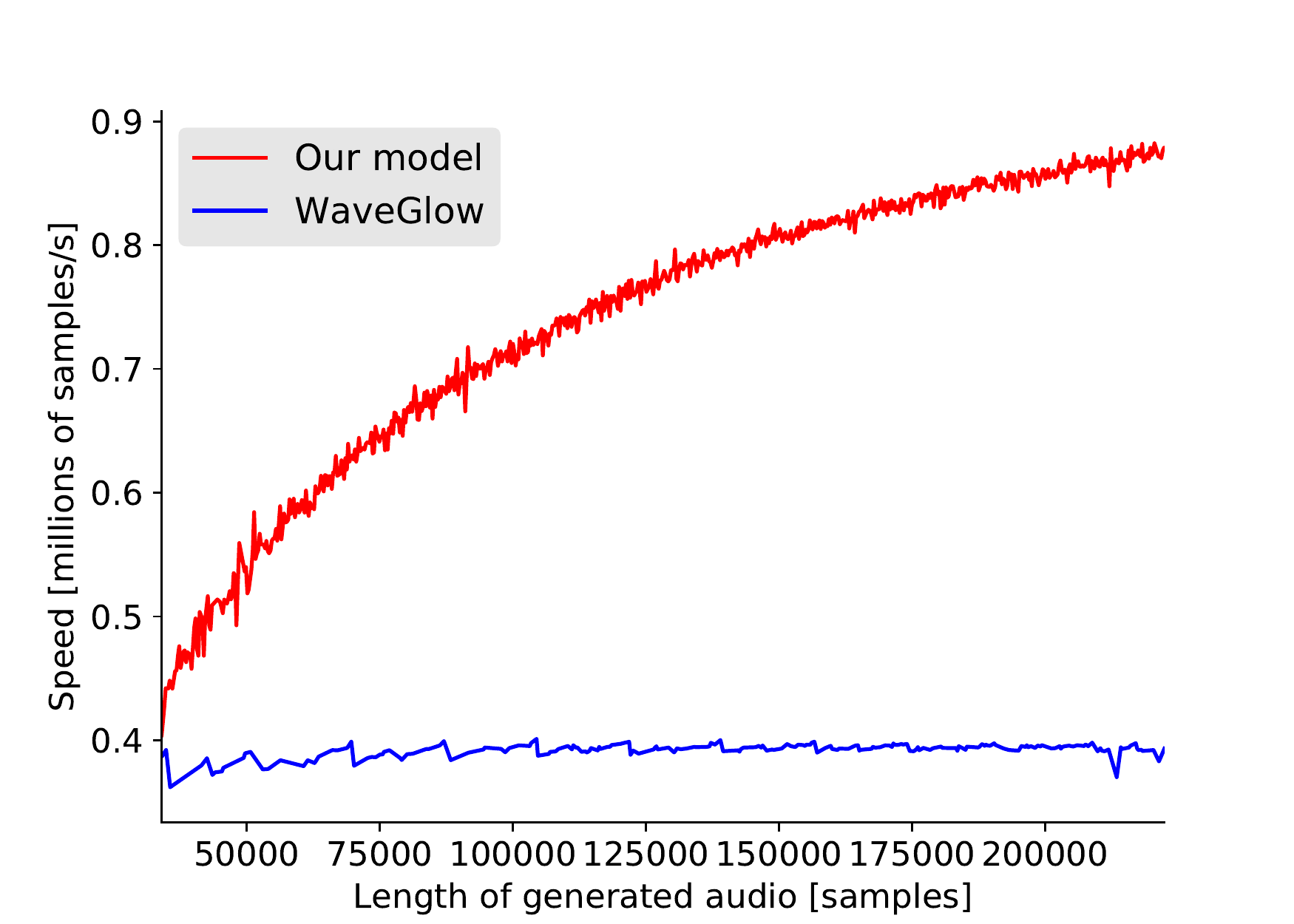}}
\end{minipage}
\caption{Synthesis speed for WaveGlow \rough{(which is fast and high-quality)} and our model. Our model can synthesize audio at almost 40$\times$ real-time for long audio sequences.
}
\label{fig:synth_time_comparison}
\end{figure}

Figure~\ref{fig:synth_time_comparison} shows a comparison of the inference speed for WaveGlow and our model, given different lengths of the audio to be synthesized. We avoid the inclusion of WaveNet, since it is a sequential approach and therefore very slow. The speed of WaveGlow is overall constant with the output size, whereas our model generates more samples per second the more samples the audio clip has, outperforming WaveGlow for clips longer than $2$~seconds. We attribute this different scaling behavior of the two models to the GPU using different optimizations for different architectures.
To confirm this, we repeated the measurement on the CPU (using a single core) and \rough{established} that the speed (in samples per second) of our model as a function of clip length is constant on the CPU.

\subsection{Training Time}\label{sec:training_time_results}
Our DDSP-based neural vocoder converged in 2.5~days on one GPU. This is relatively fast compared to other popular neural vocoders such as WaveNet, which required 4--5~days, 
or WaveGlow, which required 10~days and more than one GPU.

\subsection{Model Size}\label{sec:model_size_results}
Table~\ref{tab:size_comparison} shows the sizes of various neural vocoders, expressed as the number of learnable parameters.
\input{tables/size_comparison}
Our neural vocoder is relatively small: without compression it needs less than \SI{20}{MB} of memory space. Such a small size makes it possible to utilize our model on edge devices. Unfortunately, the majority of such devices lacks the presence of a GPU, therefore the inference speed for now would be very low. On the other hand, once that issue is solved, our neural vocoder will represent a viable solution also for embedded systems.

\subsection{Listening Test}

We performed a MUSHRA listening test~\cite{mushra,webmushra} with seven clips from the test set, which offers statistical significance for small numbers of participants.
To reject unreliable listeners who give high scores to bad audio 
as per the MUSHRA protocol,
we created bad audio
by randomly shifting in time each frequency band of ground-truth audio by up to 5 spectrogram frames, adding the elementwise product of the original signal with white noise, and low-pass filtering.
Of the 37 people who participated in our listening test, 8 were filtered out because they either assigned a score below 90 to a hidden copy of the reference, or they assigned a score higher than 30 to the lower anchor.
MUSHRA scores for our neural vocoder had a median of 40 (out of 100), with a relatively high variance
, because, even though the speaker's timbre is well modeled, not all consonants sound realistic and it is easy to tell it apart from the real audio clip.
The Griffin--Lim algorithm~\cite{griffinlim} had similar scores to our neural vocoder, even though its outputs have different characteristics, as consonants did not sound artificial but the speaker's timbre is contaminated by the imperfect estimation of the phase.
This indicates that a few-dimensional quality score cannot reflect many dimensions of audio quality, especially in the case of \rough{moderate scores}.
In other words, moderate scores do not tell us which of the dimensions of quality are worth improving.
WaveNet and WaveGlow yield results that are good along all dimensions of quality, and hence also received a high one-dimensional score, namely a median MUSHRA score of 90 and 84, respectively.

\section{Conclusions and Future Work}\label{sec:future_work}

\extrafootertext{
This work was supported by the Munich Center for Machine Learning (Grant No.\ 01IS18036B) and the BMBF project MLwin.}

This paper  proposed using DDSP operators within neural vocoders. This offers the possibility of controlling the pitch contour and other factors of variation of an utterance. Even though the quality of the synthesized speech can be improved, the control capabilities that the model offers open up new research directions to study further how speech is generated and new opportunities for many application fields. 

Future research directions include the usage of a different synthesis operator to make the model faster on the CPU: one example could be to use a wavetable oscillator instead of a harmonic one. Moreover, taking into account the dependency between pitch and timbre will improve synthesis quality.
Finally, disentangling the factors of variation from the mel spectrogram enables the usage of other statistical models together with our neural vocoder: such models would operate on the factors of variation and could provide natural-sounding variations to generate alternative versions of the same utterance.

\bibliographystyle{IEEEbib_abbrv}
\bibliography{main}

\end{document}

%% file: tables/size_comparison.tex
\begin{table}
	\centering
	\begin{tabular}{cc}
	\toprule
	Model & Size \\
	\midrule
	WaveNet & 6.8 million \\
	WaveGlow & 87.8 million \\
	Our model & 4.9 million \\
	\bottomrule
	\end{tabular}
	\caption{Numbers of parameters of different neural vocoders. Our model is much smaller than WaveGlow and is comparable to WaveNet. Moreover, our model offers control of interpretable factors of variation of speech, such as pitch.}
	\label{tab:size_comparison}
\end{table}